\documentclass[12pt, twocolumn,english]{iopart}
\expandafter\let\csname equation*\endcsname\relax
\expandafter\let\csname endequation*\endcsname\relax
\usepackage{amsmath}
\usepackage[T1]{fontenc}
\usepackage[english]{babel}
\usepackage{bm}
\usepackage{bbm}
\usepackage{iopams}
\usepackage{amssymb}
\usepackage{bbold}
\usepackage{graphicx}
\usepackage{appendix}
\usepackage{tikz}
\usepackage{caption,booktabs}
\usepackage{array}
\newcolumntype{P}[1]{>{\centering\arraybackslash}p{#1}}

\usepackage{color}

\usepackage[%
  breaklinks=true,
  colorlinks=true,
  linkcolor=blue,
  urlcolor=blue,
  citecolor=blue
]{hyperref}
 \usepackage[numbers,sort&compress]{natbib}
 \usepackage{hypernat}
 \usepackage{pifont}

\begin{document}
\title{Engineering spin-wave spectrum via the magnetization inertia tensor\\
}

\author{Subhadip Ghosh$^1$, Darpa Narayan Basu$^1$, Ritwik Mondal$^{1}$}
\address{
$^1$Department of Physics, Indian Institute of Technology (Indian School of Mines) Dhanbad, IN-826004, Dhanbad, India
}

\ead{ritwik@iitism.ac.in}

\begin{abstract}
Magnetic inertial dynamics has recently been predicted and experimentally demonstrated in two-sublattice ferromagnets such as CoFeB and NiFe permalloy. In this work, we investigate the spin-wave spectrum of such systems by incorporating the complete magnetic inertia tensor. By decomposing the tensor into symmetric and antisymmetric components, we identify isotropic, anisotropic, and chiral contributions to magnetic inertia. Within linear spin-wave theory, we find that the spectrum comprises two precessional and two inertial magnon bands. Remarkably, the upper precessional band intersects the lower inertial band within the Brillouin zone. Both cross-sublattice and chiral components of the inertia tensor act as effective control parameters for tuning these magnonic band structures. Furthermore, we show that the inertial spin-wave spectrum becomes nonreciprocal along propagation directions where the Dzyaloshinskii–Moriya interaction is finite. Strikingly, a similar nonreciprocity can also arise purely from chiral inertia, even in the absence of Dzyaloshinskii–Moriya interaction. Our findings establish magnetic inertia as a new pathway to engineer nonreciprocal magnon transport and ultrafast spintronic functionalities.
    
\end{abstract}
      
\section{Introduction}
Magnetic inertia has emerged as a key concept in understanding ultrafast magnetization processes. While the classical Landau–Lifshitz–Gilbert (LLG) equation describes magnetization dynamics with precessional and damping torque, recent theoretical and experimental studies have shown that an additional torque due to the second-order time-derivative of magnetization is required to capture inertial effects in ultrafast magnetization dynamics \cite{Ciornei2011,Suhl1998, Mondal2017Nutation,Olive2012,unikandanunni2021inertial,cherkasskii2020nutation,Kikuchi2015}. Such effects have been predicted and observed in two-sublattice ferromagnets, including CoFeB and NiFe permalloy, where they manifest as frequency-dependent corrections to spin precession and can play an essential role on femtosecond timescales \cite{neeraj2021inertial,De2025PRB}.

The idea of inertial magnetization dynamics was introduced by Ciornei {\it et al.} \cite{Ciornei2011}, who proposed augmenting the LLG equation with an inertial term to account for the finite relaxation time of angular momentum transfer. This theoretical prediction suggested the existence of nutational motion of the magnetization at sub-picosecond scales \cite{Wegrowe2012,Wegrowe2015JAP,Wegrowe2016JPCM,Titov2021Inertial,He2023}. Subsequent work by F\"ahnle and co-workers reformulated the theory in a tensorial form, identifying both isotropic and anisotropic contributions to the inertia \cite{Fahnle2011,fahnle2013erratum}. Several other theories have been proposed, including relativistic higher-order spin–orbit coupling \cite{Mondal2017,Mondal2018JPCM}, dynamical Ruderman–Kittel–Kasuya–Yosida (RKKY) exchange mediated by conduction electrons \cite{Jansen2025,Kachkachi2025}, derivations from the classical mechanics of a current-carrying loop \cite{Giordano2020}, quantum transport formalisms, such as the time-dependent non-equilibrium Green’s functions and Keldysh approach \cite{Bhattacharjee_2012,Bajpai2019}. Other mechanisms involve stochastic and dissipative origins, such as the fluctuation-dissipation theorem applied in the non-Ohmic regime \cite{Anders_2022}, or bath-induced spin inertia where phonons couple to the spin system \cite{Gaspar2024}.  Experimental verification followed, with ultrafast pump–probe and THz excitation experiments confirming the presence of nutation in materials such as CoFeB and permalloy, thereby validating the magnetic inertial dynamics and corresponding nutation resonance at the THz frequencies \cite{neeraj2021inertial,unikandanunni2021inertial,De2025PRB}. More recently, theoretical studies have extended the magnetic inertia to multi-sublattice magnets, where additional cross-sublattice inertia terms can adequately modify spin-wave spectra \cite{Mondal2021PRB,Mondal2021JPCM,CherkasskiiPRB2024,Mondal2022PRB}.

Magnetic inertial dynamics is commonly characterized by the inertial relaxation time, $\eta$, treated as a scalar parameter \cite{MONDAL_Review}. However, a significant discrepancy exists between theoretical predictions and experimental measurements of $\eta$. For instance, in polycrystalline NiFe, $\eta$ was reported to be about 300 fs \cite{neeraj2021inertial}, while other experiments have measured values as large as 1.6 ps \cite{De2025PRB}. In contrast, {\it ab initio} calculations predict much shorter timescales, typically within a few femtoseconds or even smaller for transition metals such as Fe, Co, and Ni \cite{thonig2017magnetic,Bajaj2024}. Nonetheless, it was suggested that       
the magnetic inertia is a tensor \cite{thonig2017magnetic,Bhattacharjee_2012,Ghosh2024,MONDAL_Review,Bajaj2024,Juba2019PRM}. Such a tensor can generally be decomposed into symmetric and antisymmetric parts. The symmetric part corresponds respectively to isotropic or anisotropic magnetic inertia, while the antisymmetric part corresponds to the chiral magnetic inertia \cite{Ghosh2024,Dhali_2024}. 

On the other hand, recent advances in anisotropic exchange interactions highlight the role of the Dzyaloshinskii–Moriya interaction (DMI), which arises from spin–orbit coupling and breaks inversion symmetry. This asymmetry gives rise to nonreciprocal spin-wave dispersion in ferromagnets, where the frequencies of magnons at $k$ and $-k$ differ \cite{Coldea_2002,udvardi2009,Costa_2010,dos_santos_2020,Moon2013}. Such effects were traditionally observed only for precessional magnons. More recently, however, it has been demonstrated that DMI also induces nonreciprocal dispersion in high-frequency inertial magnons. Furthermore, it has been shown that the effective gyromagnetic ratio $\gamma_{\rm eff}$, the effective damping parameter $\alpha_{\rm eff}$, and the magnon group velocity become wave-vector dependent and vary with the inertial relaxation time $\eta$ \cite{Mondal2022PRB,CherkasskiiPRB2024,Cherkasskii2021}. Despite the progress, the interplay between the magnetic inertia tensor and DMI, and the resulting modifications in magnon spectra, remains largely unexplored in two-sublattice ferromagnets. In particular, the consequences of this interplay for the magnon band structure, including possible band crossings between precessional and inertial magnon modes, have not been systematically studied.
 
In this work, we investigate the spin-wave spectrum of two-sublattice ferromagnets in the presence of a general inertia tensor and DMI. In particular, we calculate the spin-wave spectrum induced by the antisymmetric part of the inertia tensor, commonly referred to as chiral magnetic inertia.  Using linear spin-wave theory, we demonstrate the emergence of two precessional and two inertial magnon bands in two-sublattice ferromagnets on a rectangular lattice. Our results show that while the spin-wave spectrum preserves the reciprocity without DMI, such reciprocity breaks down with DMI. The latter also holds for the inertial magnon bands.  We report a band crossing between the upper precessional and lower inertial magnon bands within the Brillouin zone. We further show that these magnon precessional and inertial bands can be tuned via cross-sublattice inertia and chiral inertia.  
The cross-sublattice inertia enhances the velocity of inertial magnons compared to the precessional ones while 
simultaneously reducing their effective damping. In contrast, the chiral inertia induces an asymmetry in the 
spin-wave spectrum even in the absence of DMI, thereby providing an independent mechanism to achieve nonreciprocal 
magnon propagation.
To this end, we compute the magnon group velocities and effective damping. Our results establish a direct link between magnetic inertia and tunable magnonic band engineering, opening avenues for low-dissipation transport, nonreciprocal magnon-based devices, and ultrafast terahertz spintronics.

\section{Linear spin wave theory}\label{Sec2}
\subsection{Lattice Structure}
\begin{figure}[tbh!]
    \centering
\includegraphics[scale=0.5]{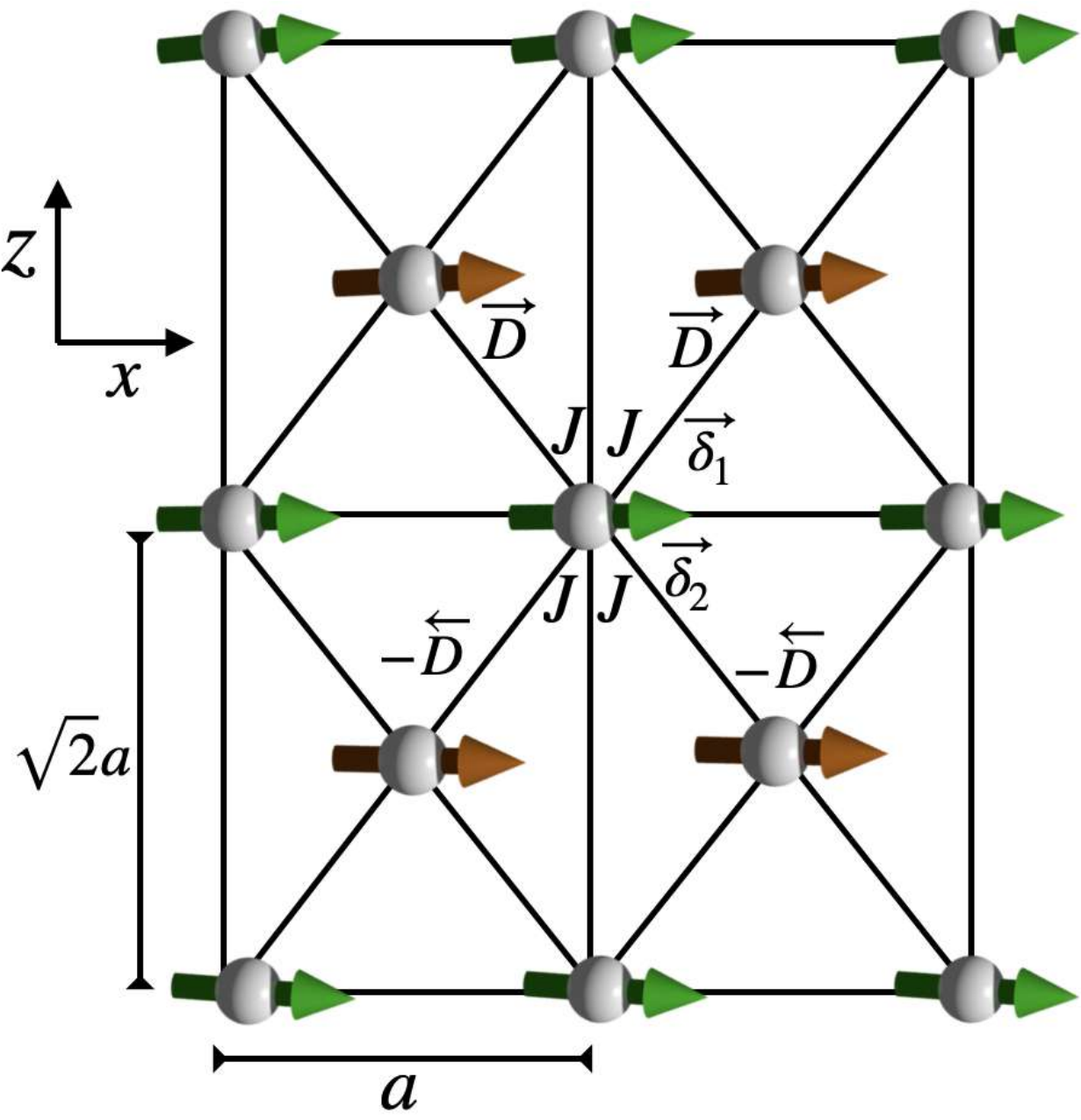}
    \caption{Sketch of a two-dimensional rectangular lattice. The ground state of the two-sublattice ferromagnet is shown in the $x$-direction with green arrows belonging to sublattice $A$ and brown arrows belonging to sublattice $B$. ${\bm\delta}_1$ and ${\bm \delta}_2$ represent the nearest-neighbor lattice vectors. The nearest neighbor exchange coupling constants are represented as $J$, with $J_{xx}$ and $J_{yy} = J_{zz}$. The Dzyaloshinskii-Moriya interaction is denoted by $\bm{D}$.}
    \label{fig:1}
\end{figure}

We consider two-sublattice ferromagnets on a rectangular lattice with lattice parameters $a$ along $x$-direction and $\sqrt{2}a$ along $z$-direction as shown in Fig. \ref{fig:1}. Two types of magnetic atoms have been represented by the two colors in Fig. \ref{fig:1}, and later, it reflected in the preceding equations as $A$ and $B$. Such a type of magnetic structure may be observed at the interface CoFeB/W or CoFeB/Ta \cite{chaurasiya2021influence,Gross2015,Torrejon2014,Soucalline2016}. It is known that such magnetic systems show interfacial DMI, and the strength depends on the thickness of the magnetic layer \cite{Chourasia2019,Kim2013,udvardi2009,Zakeri}. Further, we note that the magnetic inertial dynamics has been experimentally realised in CoFeB, showing inertial relaxation time $\eta \sim 300$ fs \cite{neeraj2021inertial}.

We restrict the Heisenberg exchange coupling to the nearest neighbor interactions i.e., $J_{xx}$ and $ J_{yy}$ = $J_{zz}$, and the DMI to the next nearest neighbor, where $D_{ij} = J_{ij}^{xy} = - J_{ij}^{yx}$. The equilibrium magnetization direction is ${\hat{\bm x}}$. The nearest neighbor distances are represented by two vectors $\bm{\delta}_1 = \frac{a}{2}\bm{\hat{x}}+\frac{a}{\sqrt{2}}\bm{\hat{z}}$ and $\bm{\delta}_2= \frac{a}{2}\bm{\hat{x}}-\frac{a}{\sqrt{2}}\bm{\hat{z}}$ as indicated in Fig. \ref{fig:1}. For such rectangular lattice system, the Brillouin zone (BZ) lies $-\pi/a \le {\rm BZ}\le \pi/a$ along the ${\hat{\bm x}}$-direction and  $-\pi/\sqrt{2}a \le {\rm BZ}\le \pi/\sqrt{2}a$ along the ${\hat{\bm z}}$-direction. We have used $a = 1$ {\AA} in the preceding calculations.

\subsection{Atomistic Spin Hamiltonian and iLLG equation}
To account for the total energy of the above system, we consider the following atomistic spin Hamiltonian
\begin{align} \label{Eq1}
    \mathcal{H} & = \frac{1}{2}\sum_{i\neq j} \bm{\sigma}^A_i \bm{J}_{ij} \bm{\sigma}^B_j + \sum_i \bm{\sigma}^A_i\bm{K}^A_i\bm{\sigma}^A_i+ \sum_j \bm{\sigma}^B_j\bm{K}^B_j\bm{\sigma}^B_j  -\sum_i \bm{B}_iM^A_i\bm{\sigma}^A_i- \sum_j \bm{B}_jM^B_j\bm{\sigma}^B_j\,,
\end{align}
where $\bm{\sigma}_i$ is the unit spin vector, $\bm{J}_{ij}$ is the exchange interaction between $i$-th and $j$-th lattice sites, $\bm{K}_i$ is the onsite magnetic anisotropy, $M_i$ is magnitude of the magnetization in site $i$ and $\bm{B}_i$ in the external magnetic field acting on the spin system. While the subscripts `$i$' denote the lattice sites, the superscripts signify $A$ and $B$ atoms.
{ The considered $\bm{J}_{ij}$ is a full 3$\times$3 matrix which can be decomposed as symmetric and antisymmetric parts. While the symmetric part accounts for isotropic Heisenberg exchange interaction, the antisymmetric part provides the DMI as $\mathcal{H}_{\mathrm{DMI}} 
    = \bm{D}_{ij} \cdot \left( \bm{\sigma}_i \times \bm{\sigma}_j \right)$, where $\bm{D}_{ij}$ is the DM vector \cite{Moriya1960}. Further, we have not considered the dipolar interaction term, assuming it is negligibly small at the short-wavelength limit, where exchange interaction dominates.}

The time evolution of the atomic spins is described by the inertial  Landau-Lifshitz-Gilbert (iLLG) equation, having the following form \cite{Osorio2024,Mondal2016,Bhattacharjee_2012,Fahnle2011,Bajaj2024}
\begin{align}
\label{Eq2}
    \frac{\partial \bm{\sigma}_i}{\partial t} & = \gamma_i\bm{\sigma}_i\times \bm{H}^{\rm eff}_i \,+ \,\frac{\alpha_i}{M_{i}}\bm{\sigma}_i\times \frac{\partial \bm{\sigma}_i}{\partial t} + \frac{1}{M_i}\bm{\sigma}_i\times \sum_j\left(\bm{\Delta}_{ij}\cdot \frac{\partial^2 \bm{\sigma}_j}{\partial t^2}\right)\,.
\end{align}
Here, $\gamma_i$ is the gyromagnetic ratio, $\alpha_i$ the Gilbert damping parameter, and $\boldsymbol{\Delta}_{ij}$ the magnetic inertia tensor. The effective field is
\begin{align}
\boldsymbol{H}^{\rm eff}_i = -\frac{1}{M_i}\frac{\partial \mathcal{H}}{\partial \boldsymbol{\sigma}_i}\,,
\label{Eq3}
\end{align}
with $\mathcal{H}$ defined in Eq.~(\ref{Eq1}). We decompose $\boldsymbol{\Delta}_{ij}$ into three parts: (i) an isotropic inertia $\eta_{ij}$; (ii) a symmetric, traceless anisotropic component $S_{ij}$; and (iii) an antisymmetric, chiral component represented by the axial vector $\boldsymbol{C}_{ij}$~\cite{Ghosh2024}. We term such chiral contribution as the chiral inertia. 
The case $j=i$ corresponds to the intra-sublattice inertia $\eta_{ii}\equiv \eta_i$, which has been discussed previously~\cite{Ciornei2011}; $j\neq i$ defines the cross-sublattice inertia $\eta_{ij}$~\cite{Mondal2021PRBSpinCurrent}. In evaluating cross-sublattice contributions we restrict to nearest neighbors. Similarly, chiral inertia exists naturally for $j\neq i$. Intra-sublattice chiral inertia  ($j=i$) would require local inversion-symmetry breaking and is generally expected to be much weaker than its cross-sublattice counterpart ${\bm C}_{ij}$. 

We also neglect the symmetric traceless tensor $S_{ij}$: its diagonal elements renormalize the isotropic inertia similarly to $\eta_i$, while its off-diagonal elements make only minor contributions. { The diagonal parts of the symmetric inertia tensor change the magnetization precessional frequency by only about 1.8\% at smaller inertial relaxation time \cite{Ghosh2024}.}  
Considering all the above facts, we recast the iLLG equations for two-sublattice system as
\begin{align}
\label{Eq4}
    \frac{\partial {\bm \sigma}^A_i}{\partial t}  = & -\gamma^A_i {\bm \sigma}^A_i \times \bm{H}^{\rm eff}_i \,+ \,\frac{\alpha^A_i}{M^A_{i}}\left(\bm{\sigma}^A_i \times \frac{\partial \bm{\sigma}^A_i}{\partial t}\right)  +\frac{\bm{\sigma}^A_i}{M^A_i}\times \left[\eta^{AA}_{ii} \frac{\partial^2 \bm{\sigma}^A_i}{\partial t^2}  
    -\left({\bm C}^{AA}_{ii}\times\frac{\partial^2 \bm{\sigma}^A_i}{\partial t^2}\right)\right]
    \nonumber\\&+\frac{\bm{\sigma}^A_i}{M^A_i} \times\left[\eta^{AB}_{ij} \frac{\partial^2 \bm{\sigma
    }^B_j}{\partial t^2} - \left(\bm{C}^{AB}_{ij}\times\frac{\partial^2 \bm{\sigma}^B_j}{\partial t^2}\right)\right]\\ 
    \frac{\partial \bm{\sigma}^B_j}{\partial t}  = & -\gamma^B_j \bm{\sigma}^B_j \times {\bm H}^{\rm eff}_j \,+ \,\frac{\alpha^B_j}{M^B_{j}}\left(\bm{\sigma}^B_j \times \frac{\partial \bm{\sigma}^B_j}{\partial t}\right)  +\frac{\bm{\sigma}^B_j}{M^B_j}\times \left[\eta^{BB}_{jj} \frac{\partial^2 \bm{\sigma}^B_j}{\partial t^2}  -\left(\bm{C}^{BB}_{jj}\times\frac{\partial^2 \bm{\sigma}^B_j}{\partial t^2}\right)\right]\nonumber\\&+\frac{\bm{\sigma}^B_j}{M^B_j} \times\left[\eta^{BA}_{ji} \frac{\partial^2 \bm{\sigma
    }^A_i}{\partial t^2} - \left(\bm{C}^{BA}_{ji}\times\frac{\partial^2 \bm{\sigma}^A_i}{\partial t^2}\right)\right]
    \label{Eq5}
\end{align}
where $\eta_{ii}^{AA}$ and $\eta_{jj}^{BB}$ are the traditional scalar inertia representing the inertial relaxation time \cite{CherkasskiiPRB2024}. Similarly, $\eta_{ij}^{AB}$ and $\eta_{ji}^{BA}$ represent the cross-sublattice scalar inertia terms \cite{Mondal2021PRB}. Note that such cross-sublattice terms have been discussed in the context of Gilbert damping \cite{Kamra2018,Sun2022}.  While ${\bm C}^{AA}_{ii}$ and  ${\bm C}^{BB}_{jj}$ are the intra-sublattice chiral inertia, $\bm{C}_{ij}^{AB}$ and $\bm{C}_{ji}^{BA}$ are the cross-sublattice chiral inertia. We have computed the effect of these intra-sublattice chiral inertia via the linear spin wave theory and their contributions are extremely small in our parameter range. Hence, in what follows, we omit the same-sublattice chiral inertia term.

In this work, we compute the spin-wave dispersion, group velocity, and the effective damping for a two-sublattice ferromagnet within the first Brillouin zone. In particular, we concentrate on the influence of the intra-sublattice inertia $\eta_i$, cross-sublattice inertia $\eta_{ij}$, and the cross-sublattice chiral inertia ${\bm C}_{ij}$. To obtain the dispersion relation, we consider $\gamma^A_i = \gamma^A$,  $\gamma^B_j = \gamma^B$, $M^A_i = M_A$, $M^B_j = M_B$, $\alpha^A_j = \alpha^A$, $\alpha^B_j = \alpha^B$, 
$\eta_{ii}^{AA} = \eta^{AA}$, $  \eta_{jj}^{BB}  = \eta^{BB}$, $\eta_{ij}^{AB} = \eta^{AB}$, $\eta_{ji}^{BA} = \eta^{BA}$, $\bm{C}_{ij}^{AB} = \bm{C}^{AB}$, $\bm{C}_{ji}^{BA}=\bm{C}^{BA}$, and a uniform external magnetic field {${\bm B}_i = {\bm B}_j = B_x\hat{\bm x}$}.

We assume that the ground state of the two-sublattice ferromagnet remains along the $\hat{\bm x}$ direction [see Fig. \ref{fig:1}]. The small deviations from the equilibrium can be expressed in terms of the angle variables $\beta_{1i}$ and $\beta_{2i}$.  The spin directions can thus be expanded as \cite{Rozsa_2013,Mondal2022PRB}
\begin{align}
\label{Eq6}
    \bm{\sigma}^A_i & = \begin{pmatrix}
    1 - \dfrac{({\beta}_{1i}^{A})^2}{2} - \dfrac{({\beta}_{2i}^{A})^2}{2}\\
    {\beta}^A_{2i}\\
    - {\beta}^A_{1i}
\end{pmatrix}\,,\\
    \bm{\sigma}^B_j & = \begin{pmatrix}
    1 - \dfrac{({\beta}_{1j}^{B})^2}{2} - \dfrac{({\beta}_{2j}^{B})^2}{2}\\
    {\beta}^B_{2j}\\
    - {\beta}^B_{1j}
    \end{pmatrix}\,.
    \label{Eq7}
\end{align}
Using these expansions, we first compute the spin Hamiltonian in Eq. (\ref{Eq1}) and consequently calculate the effective fields $\bm{H}^{\rm eff}_i$  and $\bm{H}^{\rm eff}_j$. We use these effective fields in the iLLG Eqs. (\ref{Eq4}) and  (\ref{Eq5}) and linearize those.  
We restrict ourselves to the linear terms in $\beta_{2i}$ and $\beta_{1i}$. 
Invoking a spatial Fourier transformation for the angle variables as
\begin{align}
\label{Eq8}
\tilde{\beta}_{1(2)}^{A}(\bm{k}) & = \frac{1}{\sqrt{N}} \sum_{\bm{R}_{i}^{A}} e^{-{\rm i} \bm{k}\cdot \bm{R}_{i}^{A}} \beta_{1i(2i)}^{A}\,,\\
\tilde{\beta}_{1(2)}^{B}(\bm{k}) & = \frac{1}{\sqrt{N}} \sum_{\bm{R}_{j}^{B}} e^{-{\rm i} \bm{k}\cdot \bm{R}_{j}^{B}} \beta_{1j(2j)}^{B}\,,
\label{Eq9}
\end{align}
and moving to the circular basis defined by $\tilde{\beta}_\pm^A(\bm{k}) = \tilde{\beta}_2^A(
 \bm{k})\pm {\rm i} \tilde{\beta}_1^A({\bm{k}})$ and $\tilde{\beta}_\pm^B(\bm{k}) = \tilde{\beta}_2^B(
 \bm{k})\pm {\rm i} \tilde{\beta}_1^B({\bm{k}})$, Eqs. (\ref{Eq4}) and (\ref{Eq5}) can be recast as
 \begin{align}
 \label{Eq10}
\frac{\partial\tilde{\beta}_\pm ^A({\bm k})}{\partial t} & =\frac{\gamma_A}{M_A}\biggl[ \pm{\rm i}\Omega_A\tilde{\beta}_\pm^A(\bm{k}) \mp{\rm i} \tilde{J}(\bm{k})\tilde{\beta}_\pm^B(\bm{k})+{\rm i} \tilde{D}(\bm{k})\tilde{\beta}_\pm^B(\bm{k})\biggr]  \mp{\rm i} \alpha^A \frac{\partial \tilde{\beta}^A_\pm (\bm{k})}{\partial t} \nonumber\\&\mp {\rm i} \eta^{AA} \frac{\partial ^2\tilde{\beta}^A_\pm (\bm{k})}{\partial t^2}\mp {\rm i} \tilde{\Delta}^{AB}(\bm{k}) \frac{\partial^2\tilde{\beta}^A_\pm (\bm{k})}{\partial t^2}\,, \\
    \frac{\partial \tilde{\beta}_\pm ^B(\bm{k})}{\partial t} & = \frac{\gamma_B}{M_B}\biggl[\pm{\rm i} \Omega_B\tilde{\beta}^B_\pm (\bm{k}) \mp {\rm i}\tilde{J}(\bm{k}) \tilde{\beta}_\pm^A(\bm{k})+ {\rm i} \tilde{D}(\bm{k})\tilde{\beta}_\pm^A(\bm{k})  \biggr] \mp{\rm i} \alpha^B \frac{\partial \tilde{\beta}^B_\pm (\bm{k})}{\partial t} \nonumber\\&\mp {\rm i} \eta^{BB} \frac{\partial ^2\tilde{\beta}^B_\pm (\bm{k})}{\partial t^2}\mp {\rm i} \tilde{\Delta}^{BA}(\bm{k})\frac{\partial^2\tilde{\beta}^B_\pm ({\bm k})}{\partial t^2}\,,
    \label{Eq11}
\end{align}
with the following definitions    $\tilde{\Delta}^{AB}(\bm{k})=\left(\tilde{\eta}^{AB}(\bm{k})\pm{\rm i}\tilde{C}^{AB}(\bm{k})\right)$,  $\tilde{\Delta}^{BA}(\bm{k})=\left(\tilde{\eta}^{BA}(\bm{k})\pm{\rm i}\tilde{C}^{BA}(\bm{k})\right)$, where 
\begin{align*}
\Omega_{A} & = \frac{\gamma_A}{M_A}\left(4J_{xx}+2K^A_x+B_xM_A\right)\,,\\
\Omega_B & = \frac{\gamma_B}{M_B}\left(4J_{xx}+2K^B_x+B_xM_B\right)\,,\\
    \tilde{\eta}^{AB}(\bm{k}) & =2\eta^{AB}\left[\cos(\bm{k}\cdot\bm{\delta}_1)+\cos(\bm{k}\cdot\bm{\delta}_2)\right]\,,\\
    \tilde{C}^{AB}(\bm{k}) & =2{\rm i}C^{AB}\left[\sin(\bm{k}\cdot\bm{\delta}_1)+\sin(\bm{k}\cdot\bm{\delta}_2)\right]\,.
\end{align*}  
Within the linear 
regime, we assume a harmonic time dependence of the angle variable  $\tilde{\beta}_\pm \propto e^{\pm i\omega t}$. This allows us to replace $\frac{\partial}{\partial t}$ by $\pm{\rm i}\omega$ and $\frac{\partial^2}{\partial t^2}$ by $-\omega^2$ in the Eqs. (\ref{Eq10}) and  (\ref{Eq11}) and rewrite them in matrix form as
\begin{align}\label{18}
\begin{pmatrix}
-\eta^{AA}\omega^2 \pm {\rm i}\alpha^A\omega +\omega - \dfrac{\gamma_A\Omega_A}{M_A} & \dfrac{\gamma_A\left(\tilde{J}(\bm{k})  \mp \tilde{D}(\bm{k})\right)}{M_A}-\tilde{\Delta}^{AB}_\pm(\bm{k})\omega^2\\
\dfrac{\gamma_B\left(\tilde{J}(\bm{k})  \mp \tilde{D}(\bm{k})\right)}{M_B}-\tilde{\Delta}^{BA}_\pm(\bm{k})\omega^2 & -\eta^{BB}\omega^2 \pm {\rm i}\alpha^B\omega +\omega - \dfrac{\gamma_B\Omega_B}{M_B}
\end{pmatrix}
\begin{pmatrix} \tilde{\beta}^A_\pm (\bm{k})\\
\tilde{\beta}^B_\pm (\bm{k}) 
\end{pmatrix} & = \begin{pmatrix}  0\\
 0
\end{pmatrix}.
\end{align}
 
\subsection{Dispersion relation}

To obtain the dispersion relation, we set the determinant of the corresponding matrix in Eq. (\ref{18}) to zero. Such equation has the following form
\begin{align}
    \mathcal{P}_{\pm}\omega^4(\bm{k})+\mathcal{Q}_{\pm}\omega^3(\bm{k})+\mathcal{R}_{\pm}\omega^2(\bm{k})+\mathcal{S}_{\pm}\omega(\bm{k})+\mathcal{T}_{\pm}(\bm{k})=0\,, \label{Eq12}
\end{align}
where the coefficients are
\begin{align}
    \label{Eq13}
\mathcal{P}_{\pm}& = {\eta}^{AA}{\eta}^{BB}-\tilde{\Delta}^{AB}(\bm{k})\tilde{\Delta}^{BA}(\bm{k})\,,\\
    \label{Eq14}\mathcal{Q}_{\pm} & = -\left(\eta^{AA}+\eta^{BB}\right)\mp{\rm i}\left(\alpha_B\eta^{AA}+\alpha_A\eta^{BB}\right)\,,\\
    \label{Eq15}\mathcal{R}_{\pm} & = 1-\alpha_A\alpha_B +\left(\Omega_A\eta^{BB}+\Omega_B\eta^{AA}\right)\pm {\rm i}\left(\alpha_A+\alpha_B\right)\nonumber\\&+\left(\tilde{J}(\bm{k})\pm\tilde{D}(\bm{k})\right)\left(\frac{\gamma_A\tilde{\Delta}^{BA}}{M_A}+\frac{\gamma_B\tilde{\Delta}^{AB}}{M_B}\right)\,,\\
\label{Eq16}    \mathcal{S}_{\pm}& = -\Omega_A-\Omega_B\mp{\rm i}\left(\alpha_A\Omega_B+\alpha_B\Omega_A\right)\,,\\
    \label{Eq17}\mathcal{T}_{\pm}& =  \Omega_A\Omega_B-\frac{\gamma_A\gamma_B\left(\tilde{J}(\bm{k})\pm\tilde{D}(\bm{k})\right)^2}{M_AM_B}\,,
    \end{align}
with the following expressions
    \begin{align}
    \tilde{J}(\bm{k}) & = 2 J_{zz} \left[\cos (\bm{k}\cdot \bm{\delta}_1) + \cos (\bm{k}\cdot \bm{\delta}_2)\right]\,,\\
    \tilde{D}(\bm{k}) & = 2D \left[\sin (\bm{k}\cdot \bm{\delta}_1) - \sin (\bm{k}\cdot \bm{\delta}_2)\right]\,.\label{17}
\end{align}
The solution of Eq. (\ref{Eq12}) provides the spin-wave dispersion relation. The analytical solution of the above equation is rather cumbersome; therefore, we adopt a numerical solution. Note that Eq. (\ref{Eq12}) is a fourth-order polynomial in $\omega(\bm{k})$. Therefore, there will be four distinct solutions -- two of the solutions will signify the precessional modes, while the other two specify the inertial modes. We denote the precessional resonance frequencies as $\omega_p$ and the inertial resonance frequencies by $\omega_n$. Furthermore, to distinguish the two precessional resonance frequencies, we denote $\omega_p^l$ for the one having a lower frequency value and $\omega_p^u$ for the other having a higher frequency value, respectively. Similar notations have been used for the inertial resonance frequencies as well, $\omega_n^l$ and $\omega_n^u$. 
The above resonance frequencies are complex, having real and imaginary parts. 
We mention that there are two equations involved in Eq. (\ref{Eq12}): (a) $\mathcal{P}_{+}\omega^4(\bm{k})+\mathcal{Q}_{+}\omega^3(\bm{k})+\mathcal{R}_{+}\omega^2(\bm{k})+\mathcal{S}_{+}\omega(\bm{k})+\mathcal{T}_{+}(\bm{k})=0$ and (b) $\mathcal{P}_{-}\omega^4(\bm{k})+\mathcal{Q}_{-}\omega^3(\bm{k})+\mathcal{R}_{-}\omega^2(\bm{k})+\mathcal{S}_{-}\omega(\bm{k})+\mathcal{T}_{-}(\bm{k})=0$. While these two equations produce conjugate pair of solutions, a broken symmetry of the system (e.g., broken inversion symmetry) can lead to distinct resonance frequencies and asymmetric dispersion relation \cite{Mondal2022PRB}.

\section{Results \& discussions}
\label{Sec3}
We numerically solve Eq. (\ref{Eq12}) for a two-sublattice ferromagnet with DMI and investigate the spin-wave spectrum, group velocity, effective damping, and {effective gyromagnetic ratio}. We have considered the following assumptions: $K_x^A = K_x^B = K_x$, $\alpha_A = \alpha_B = \alpha$, $\eta^{AA} = \eta^{BB} = \eta$, $\eta^{AB} = \eta^{BA} = \eta^\prime$, $C^{AB} = C^{BA} = C^\prime$. 

\subsection{ Effect of DMI on the spin-wave spectrum with unequal magnetic moments on both sublattices}

\begin{figure}[h]
    \centering
    \includegraphics[scale=0.5]{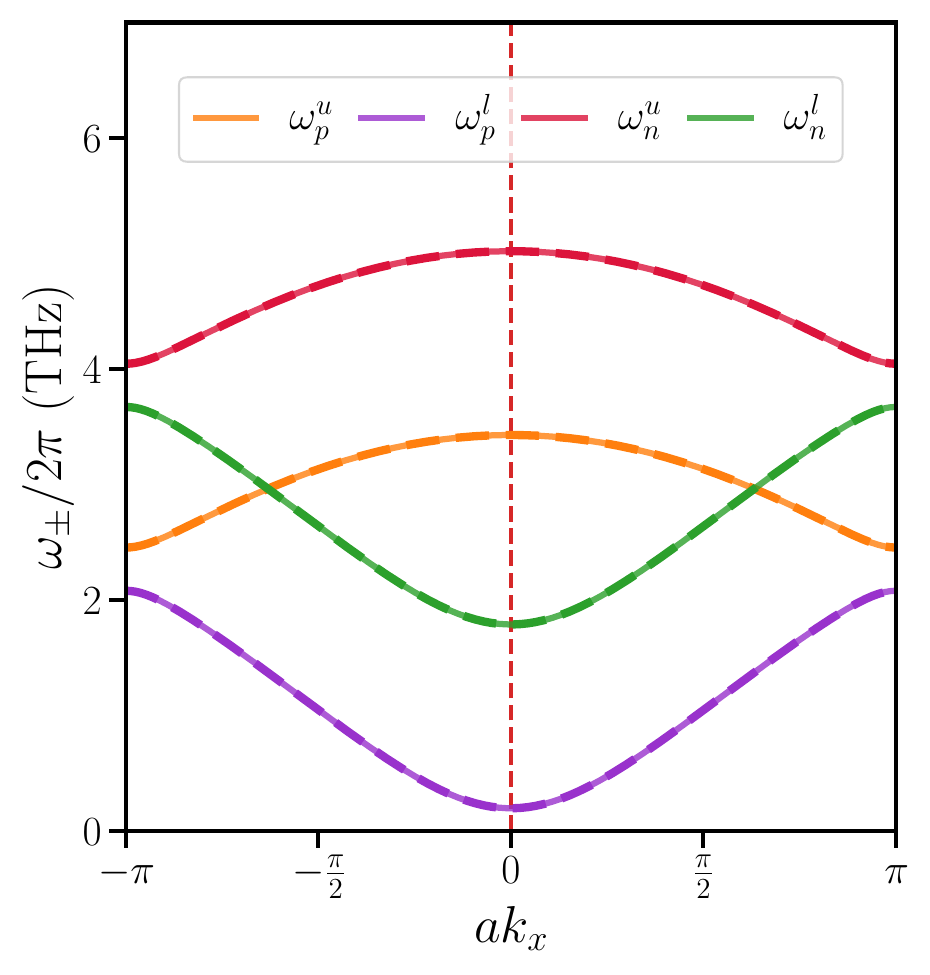}
        \caption{{ Spin-wave spectrum along the $k_x$ obtained from Eq. (\ref{Eq12}), setting {$k_z = 0$.} {The solid and dashed lines represent + and - branches of the dispersion relation.}  The other parameters are $\eta = 100$ fs, $\eta^\prime = 0$, $C^\prime = 0$, and $M_A = 2\,\,\mu_B$, $M_B = 2.6$ $\mu_B$.}}
    \label{fig:2}
\end{figure}

\begin{figure}[tbh!]
    \centering
    \includegraphics[scale=0.5]{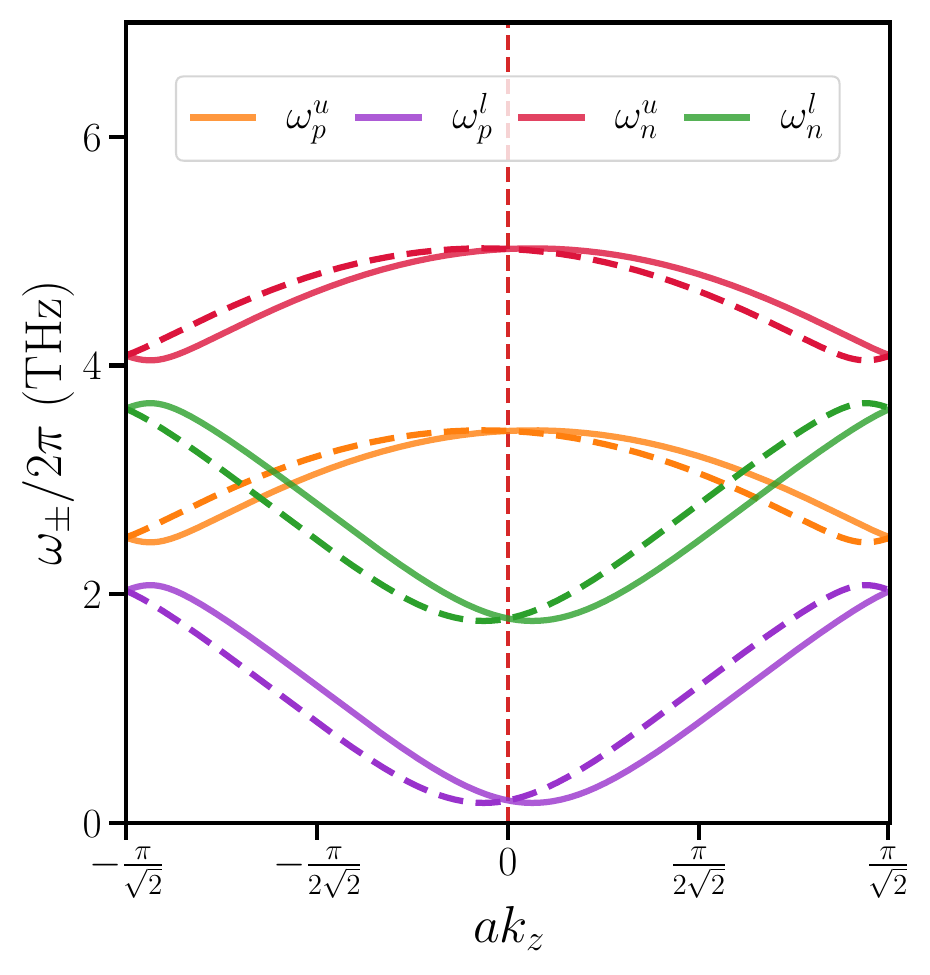}
        \caption{{ Spin-wave spectrum along $k_z$ obtained from Eq. (\ref{Eq12}), setting {$k_x = 0$.} {The solid and dashed lines represent + and - branches of the dispersion relation.} The other parameters are $\eta = 100$ fs, $\eta^\prime = 0$, $C^\prime = 0$, and $M_A = 2\,\,\mu_B$, $M_B = 2.6$ $\mu_B$.}}
    \label{fig:3}
\end{figure}

The obtained spin-wave spectrum via the solution of Eq. (\ref{Eq12}) for both the signs '+' and '-' are shown in {Fig. \ref{fig:2} for $k_z = 0$ and Fig. \ref{fig:3} for $k_x = 0$}. The parameters used are specified in Table \ref{tab:1}. In addition, we set the Gilbert damping parameter $\alpha = 0.05$ and the magnetic field $B_x = 0$ T. We observe two precessional magnon bands (purple and orange) along with two inertial magnon bands (green and red) in Fig. \ref{fig:2}. In a two-sublattice ferromagnet, the two precessional resonance frequencies follow the same rotational handedness, in contrast to the antiferromagnet, where two precessional resonances show opposite rotational sense due to the antiparallel alignment of the two sublattices \cite{Mondal2021JPCM,Mondal2020nutation}. These two precessional resonances can be observed as $\omega_{p}^l$ and $\omega_{p}^u$ in Fig. \ref{fig:2}. The corresponding inertial resonance modes are shown $\omega_{n}^l$ and $\omega_{n}^u$, respectively. {However, due to the opposite handedness, the precessional frequencies are $-\omega_p^u, -\omega_p^l$ and the corresponding inertial frequencies are $+\omega_n^u, +\omega_n^l$, as obtained by solving Eq. (\ref{Eq12}) with + sign.}
This agrees with the previously reported results, where it was shown that the precessional and inertial resonance modes have opposite handedness \cite{Kikuchi2015}.
\begin{table}[tbh!]
    \caption{Model parameters used in obtaining the numerical solutions of Eq.~(\ref{Eq12}).}
   \centering
    \begin{tabular}{ c | c | c | c }
    \hline
    \hline 
    
         $J_{xx}$ & $J_{yy}=J_{zz}$  & $D$  & $K_x$   \\
         
        \hline 
         $-1.02\times10^{-21}$ J &  $-0.99\times10^{-21} $ J & $10^{-22}$ J  &  $-10^{-23}$ J    \\
         \hline\hline
    \end{tabular}
    \label{tab:1}
\end{table}
{We plot all these frequencies by taking the absolute value.}
In the precessional magnon spectrum, the lower branch 
$\omega^{l}_{p}$ grows approximately linearly with momentum 
away from the Brillouin zone center, characteristic of an 
acoustic-like mode. In contrast, the upper branch 
$\omega^{u}_{p}$ decreases with increasing momentum, exhibiting features 
reminiscent of an optical-like mode. A similar behaviour is 
also observed in the inertial magnon branches.

As illustrated in Fig.~\ref{fig:2}, the absence of DMI restores reciprocity in the spin-wave spectrum along the 
$\hat{\bm{x}}$-direction. Consequently, the positive and negative chiral branches overlap with each other, satisfying the symmetry relation $\omega_{p}^{l}({k}_x) = \omega_{p}^{l}(-{k}_x)$
meaning the two magnon branches become energetically degenerate. Such reciprocity is preserved for other magnon branches as well along the $\hat{\bm{x}}$-direction [see Fig. \ref{fig:2}]. 
This behavior arises because the symmetric (Heisenberg) exchange interactions are invariant under momentum reversal. However, due to the 
antisymmetric exchange interaction, the effective DMI along the ${\hat{\bm z}}$-direction, such degeneracy is lifted. Therefore, it introduces a chiral asymmetry into the magnon spectrum \cite{Szulc2021,Heins2025,Di2014APL,Gubbiotti2022}. The resulting  
contribution shifts the dispersion differently for positive and negative $k_z$, giving rise to nonreciprocal propagation 
and directional anisotropy of the spin waves in the individual magnon branches. Such non-reciprocal spin-wave propagation has been shown in Fig. \ref{fig:3}.
We note that such nonreciprocity is not limited to conventional precessional spin waves only, but can also be observed in inertial spin waves. Our results agree with the previously reported spin-wave spectrum with magnetic inertia in antiferromagnets and spin spirals \cite{Mondal2022PRB,CherkasskiiPRB2024}. We further note that the nonreciprocal magnon branches are connected via the reciprocal symmetry such that $\omega_{p+}(k_z) = \omega_{p-}(-k_z)$, and $\omega_{n+}(k_z) = \omega_{n-}(-k_z)$ for upper and lower modes. Such reciprocal symmetry holds for finite Gilbert damping as $\alpha \neq 0$ in our case.  

In the spin-wave spectrum, a characteristic crossing occurs between the upper precessional magnon band ($\omega_p^u$) and the lower 
inertial magnon band ($\omega_n^l$). This crossing arises from the coexistence of conventional precessional magnetization dynamics with the magnetic
inertia-driven nutational modes, and it marks the frequency regime where their dispersions intersect. At the crossing point, the eigenmodes of the upper precessional band and the lower inertial band are opposite in character. Since experimental probes such as Brillouin light scattering or time-resolved magneto-optical 
spectroscopy is sensitive to the polarization of spin-wave modes; this difference in rotation sense offers a direct 
route to distinguish the precessional and inertial motion experimentally at the same frequency. Further, we note that $\omega_n^l$ and $\omega_p^u$ do not correspond to each other. While $\omega_n^l$ corresponds to the precessional frequencies $\omega_p^l$, the other inertial magnon branch $\omega_n^u$ corresponds to $\omega_p^u$. 

\subsection{Spin-wave spectrum with different magnetic moments on both sublattices}
\begin{figure*}[tbh!]
    \centering
    \includegraphics[scale =0.5]{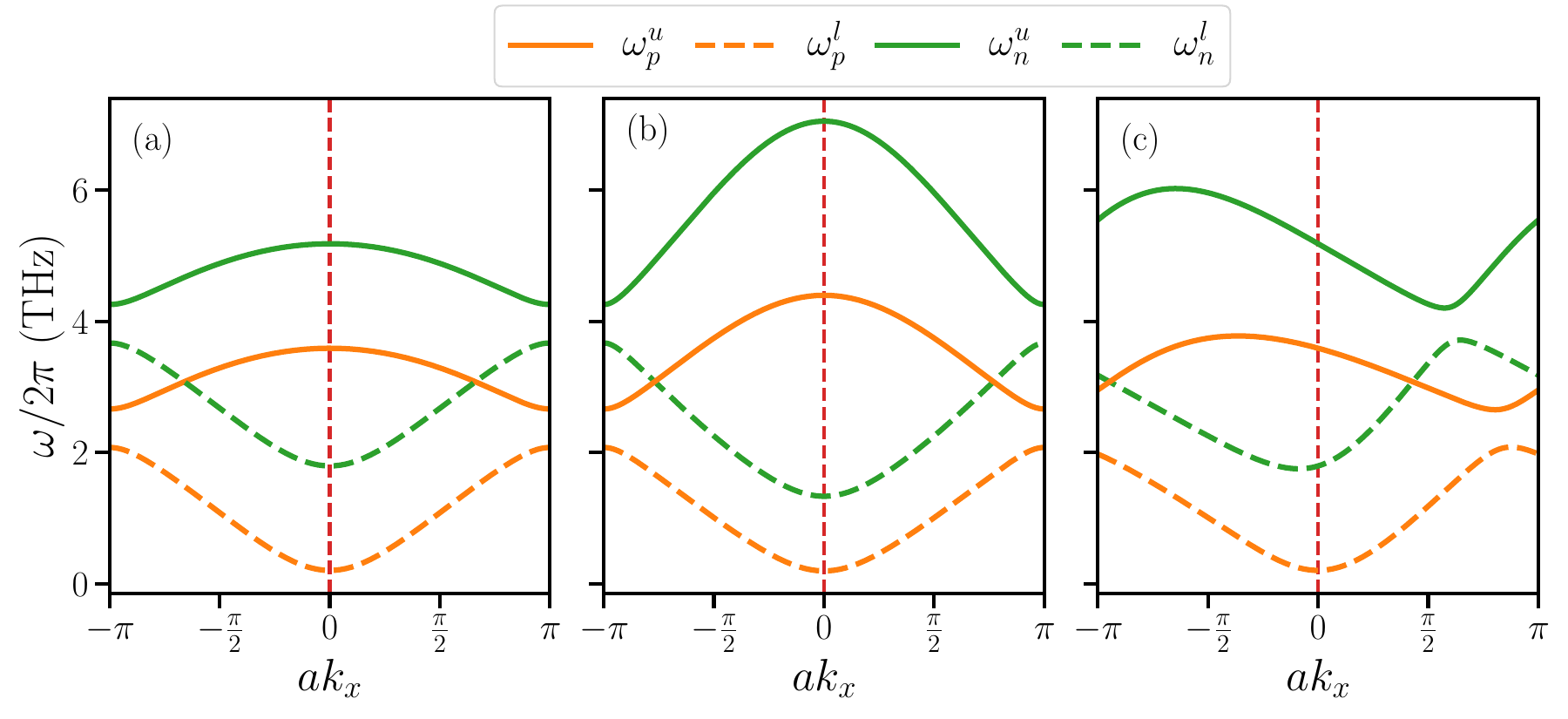}
    
    \caption{Spin-wave dispersion along {$k_x$, setting $k_z = 0$ }obtained via solution of Eq. (\ref{Eq12}). The following inertial parameters have been used (a) $\eta = 100$ fs, $\eta^\prime = 0$ fs, $C^\prime = 0$ fs (b) $\eta = 100$ fs, $\eta^\prime = 10$ fs, $C^\prime = 0$ fs, and (c) $\eta = 100$ fs, $\eta^\prime = 0$ fs, $C^\prime = 10$ fs. The other parameters are in Table \ref{tab:1}.
    }
\label{fig:4}
 \end{figure*}
In the preceding section, we discuss the spin-wave spectrum with $M_A = 1.75\, \mu_B$ and $M_B = 2.6\, \mu_B$ resembling the Co and Fe atoms in CoFeB  \cite{Frota2000,Hofherr2018,Yamamoto_2019,von_korff2023PRR}. These values provide a realistic parametrization of the two magnetic sublattices and allow us to capture the essential features of the spin-wave spectrum related to CoFeB \cite{Rana2019,Chaurasiya2016,CHOI2020,Yu2012}. The related spin-wave dispersion is shown in Fig. \ref{fig:4}. { We point out that the ‘+’ and ‘–’ branches remain degenerate in this case, coinciding along the direction where the effective DMI is absent. Thus, we have presented the results only for the ‘+’ branch.} While the dispersion relation qualitatively agrees with Fig. \ref{fig:2}, we observe that two distinct magnetic moments introduce an energy gap at the edge of the Brillouin zone for precessional magnons. It is important to note that identical magnetic moments on the two sublattices do not lead to an energy gap at the 
edge of the Brillouin zone, since the spin-wave branches remain degenerate in this case [see Fig. \ref{fig:2}]. In contrast, when the 
sublattices carry different magnetic moments, a finite gap opens at the zone boundary. This gap originates from the broken sublattice symmetry, which lifts the degeneracy between the magnon modes. Similar results can also be observed for the inertial magnons as well. We compute the energy gap as a function of the sublattice magnetization ratio ($M_B/M_A$) as well as $\eta$ in \ref{appendixA}. 
Nonetheless, the crossing between the upper precessional mode and the lower inertial mode remains clearly visible in the spin-wave spectrum.

The spin-wave spectrum in Fig. \ref{fig:4}(a) involves magnetic inertia as a scalar quantity. In fact, we have used $\eta = 100$ fs, $\eta^\prime = 0$ fs, and $C^\prime = 0$ fs. Due to the introduction of an energy gap at the edges of the Brillouin zone, the linear increase of both precessional 
and inertial frequencies away from the Brillouin zone center exhibits a reduced slope as compared to Fig.~\ref{fig:2}. However, in two-sublattice ferromagnets the precessional and corresponding inertial magnon bands follow a similar 
behavior and exhibit comparable curvature in the spin-wave spectrum. This is in stark contrast to the case of 
two-sublattice antiferromagnets, where the inertial frequencies become smaller for higher precessional frequencies 
as one moves away from the center of the Brillouin zone \cite{Mondal2022PRB}. Such a distinction underscores the fundamentally different 
role of inertial dynamics in ferromagnetic and antiferromagnetic systems.

In Fig. \ref{fig:4}(b) we explore the effect of cross-sublattice inertia $(\eta^\prime \neq 0)$ on the spin-wave spectrum. 
We find that the frequencies of the upper precessional mode ($\omega^u_{p}$) 
and the upper inertial mode ($\omega^u_{n}$) increases and attains a higher value, while the other inertial mode $\omega^l_{n}$ pushes to a rather lower value at the center of the  Brillouin zone in comparison with Fig. \ref{fig:4}(a). In contrast, the lower precessional mode 
($\omega^l_{p}$) remains almost unaffected. These trends are consistent with 
the findings of Ref.~\cite{Ghosh2024}. Further, the crossing of $\omega^u_{p}$ and $\omega^l_{n}$
occurs at a higher $k$-value within the Brillouin zone compared to Fig. \ref{fig:4}(a). 

The effect of chiral inertia has been plotted in Fig. \ref{fig:4}(c). The spin-wave spectrum becomes asymmetric with respect to the centre of the Brillouin zone. Such asymmetry is more visible in the two inertial magnon bands ($\omega^l_{n}$ and $\omega^u_{n}$). The minima of each magnon band do not correspond to $k_z = 0$, except $\omega^l_{p}$. In fact, the higher the frequency, the further its minima are shifted away from $k_z = 0$. Note that the crossing between the upper precessional band and the lower inertial band is shifted closer to the 
edge of the Brillouin zone. This asymmetry arises from the presence of chiral inertia along the direction in which 
the DMI is absent.        

The crossing of the precessional and inertial magnon bands depends strongly on the inertial relaxation time of the system. The results presented in Fig.~\ref{fig:4} used $\eta = 100$ fs. A smaller value of $\eta = 10$ fs has been employed, and the corresponding spin-wave spectrum is discussed in \ref{appendixB}. In such case, the inertial resonance frequencies follow $1/\eta$ and lie far above the precessional resonances without any overlapping frequencies.

\subsection{Magnon group velocity with different magnetic moments on both sublattices}
\begin{figure}[tbh!]
    \centering
    \includegraphics[scale = 0.5]{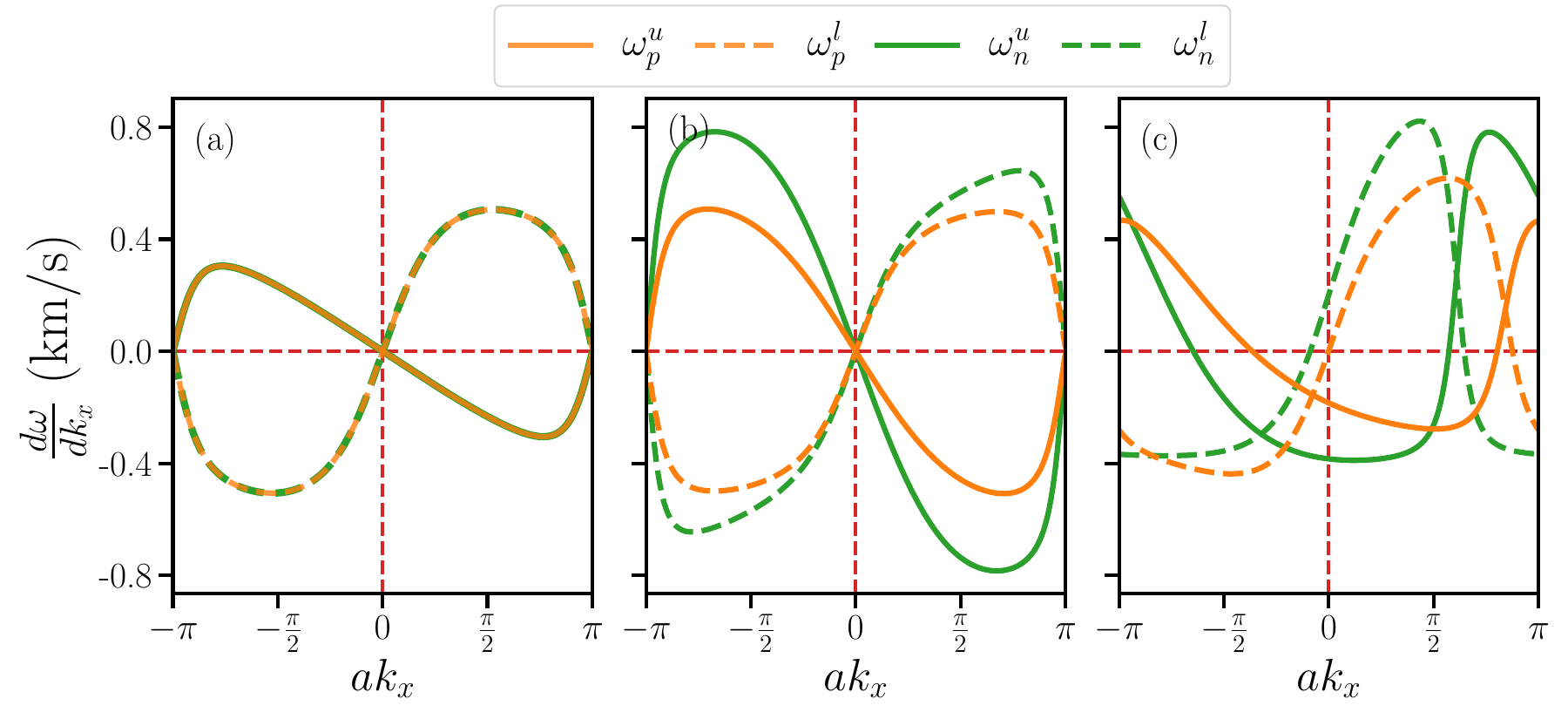}
    \caption{Magnon group velocity along $k_x$, setting $k_z = 0$ obtained via solution of Eq. (\ref{Eq12}). The following inertial parameters have been used (a) $\eta = 100$ fs, $\eta^\prime = 0$ fs, $C^\prime = 0$ fs (b) $\eta = 100$ fs, $\eta^\prime = 10$ fs, $C^\prime = 0$ fs, and (c) $\eta = 100$ fs, $\eta^\prime = 0$ fs, $C^\prime = 10$ fs. The other parameters are summarized in Table \ref{tab:1}.}
    \label{fig:5}
\end{figure}
The group velocity can directly be calculated from the spin-wave spectrum in Fig. \ref{fig:4}. It follows the expression
\begin{align}
    {\bm v}(\bm{k}) = \frac{d\omega(\bm{k})}{d\bm{k}}\,.
\end{align}
Such group velocity is presented for precessional and inertial modes in Fig. \ref{fig:5}. The group velocity vanishes at the extreme of the spin-wave spectrum e.g., at ${\bm k} = 0$ and the edge of the Brillouin zone. Although we show the group velocity along ${\hat{\bm x}}$-direction where the DMI is absent in Fig. \ref{fig:5}, the vanishing group velocity will be shifted along $k_z$ direction due to the presence of DMI \cite{Mondal2022PRB}.    

In Fig.~\ref{fig:5}(a), we investigate the effect of 
$\eta = 100$~fs, while assuming $\eta^\prime = C^\prime = 0$ fs. 
The upper precessional and inertial magnon branches exhibit 
a linear dependence on $k_{x}$, whereas the lower-frequency 
branches display a nonlinear behavior. Furthermore, the 
precessional magnons and their corresponding inertial modes 
are found to share the same group velocity. Such equality in group velocity is broken in the presence of cross-sublattice inertia ($\eta^\prime \neq 0$). Considering $\eta = 100$~fs and $\eta^\prime = 10$~fs, the calculated group velocities are shown in Fig.~\ref{fig:5}(b). We find that the group velocities of all modes are enhanced by the inclusion of cross-sublattice inertia. Moreover, the group velocities of the corresponding modes 
$\omega^u_{p}$ and $\omega^u_{n}$ are no longer identical, except at $k_x = 0$ and at the Brillouin zone edge. 
A similar trend is observed for the lower branches $\omega^l_{p}$ and $\omega^l_{n}$. 
Most importantly, the inertial magnon modes consistently exhibit higher group velocities than the corresponding 
precessional modes across the entire Brillouin zone. 
This indicates that cross-sublattice inertia-mediated spin dynamics can substantially accelerate information transport in a two-sublattice  
ferromagnetic system.

Fig. \ref{fig:5}(c) describes the effect of the chiral inertia $C^\prime = 10$ fs on the group velocity of magnon branches. Due to the asymmetric spin-wave dispersion in Fig. \ref{fig:4}(c), the vanishing group velocity does not occur at $ak_x = 0$ and $ak_x = \pm \pi/\sqrt{2}$. Further, the group velocity for precession and inertia modes becomes asymmetric with respect to $k_x = 0$ such that the propagation of spin waves is enhanced along one direction while suppressed in the opposite.

\subsection{Effective damping with different magnetic moments on both sublattices}
 \begin{figure*}[htb]
    \centering
    \includegraphics[scale=0.52]{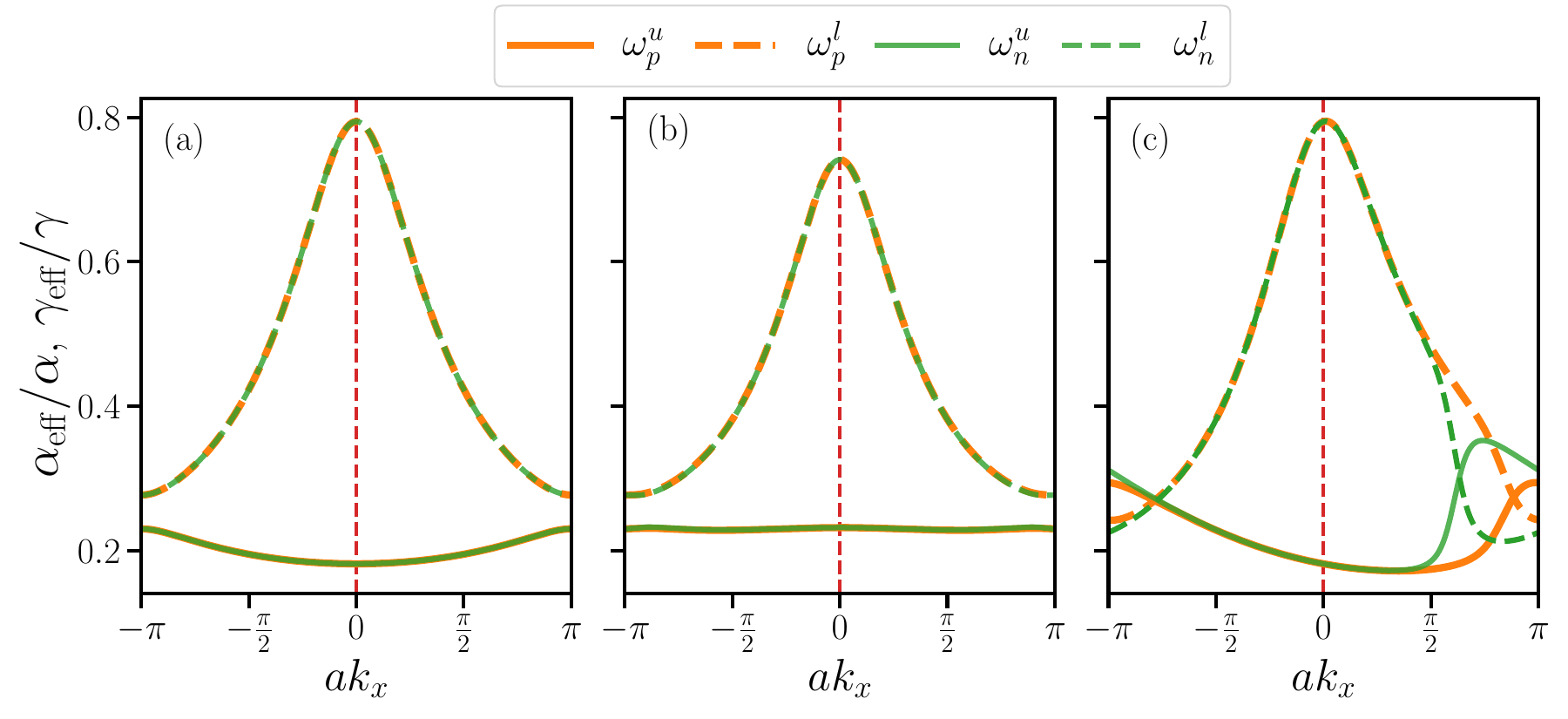}
    
    \caption{Effective damping along {$k_x$, setting $k_z = 0$} obtained via solution of Eq. (\ref{Eq12}). The following inertial parameters have been used (a) $\eta = 100$ fs, $\eta^\prime = 0$ fs, $C^\prime = 0$ fs (b) $\eta = 100$ fs, $\eta^\prime = 10$ fs, $C^\prime = 0$ fs, and (c) $\eta = 100$ fs, $\eta^\prime = 0$ fs, $C^\prime = 10$ fs. The other parameters are in Table \ref{tab:1}\,. A nonzero damping value of $\alpha = 0.05$ has been considered.}
    \label{fig:6}
\end{figure*}
Next, we compute the effective damping via
\begin{align}
    \alpha_{\rm eff}(\bm{k}) = \frac{{\tt Im}\,\left[\omega(\bm{k})\right]}{{\tt Re} \,\left[\omega(\bm{k})\right]}\,.
\end{align}
The calculated effective damping for the precessional and inertial modes is shown in Fig.~\ref{fig:6}. Since the effective gyromagnetic ratio $\gamma_{\rm eff}(\bm{k})$ follows the identical wave-vector dependence as the effective damping, following $\gamma_{\rm eff}(\bm{k}) = d\omega(\bm{k})/dB_x$, Fig.~\ref{fig:6} simultaneously illustrates the effective gyromagnetic ratio as well \cite{Mondal2022PRB}.  
Throughout the Brillouin zone, $\alpha_{\rm eff}$ for a precessional mode and its corresponding inertial partner 
completely overlap, consistent with the previously reported effective damping at $\bm{k} = 0$~\cite{Mondal2021JPCM}. 
Moving away from the zone center, the lower precessional frequency mode (together with its inertial counterpart) 
exhibits a pronounced decrease in effective damping, whereas the upper precessional frequency mode (and its inertial 
partner) displays a gradual increase. Nonetheless, the overall effective damping profile remains symmetric with 
respect to $k_x = 0$, 
reflecting the underlying symmetry of the spin-wave spectrum.   

In Fig.~\ref{fig:6}(a), we show the effect of identical sublattice inertia with $\eta = 100$~fs. 
The ratio $\alpha_{\rm eff}/\alpha$ deviates from unity because the effective damping is already reduced in the 
presence of magnetic inertia~\cite{Mondal2020nutation}. This reduction indicates a lower-dissipation 
channel mediated by inertia, which can significantly influence the lifetime and propagation length of not only the precessional magnons, but also the inertial magnons.
In Fig.~\ref{fig:6}(b), we present the effect of cross-sublattice inertia on the effective damping. At $\bm{k} = 0$, 
it is known that cross-sublattice inertia enhances the dissipation of the $\omega_{p}^u$ and $\omega_{n}^u$ modes, 
while simultaneously reducing the damping of the other two magnon modes~\cite{Mondal2021JPCM}. Our results in 
Fig.~\ref{fig:6}(b) is fully consistent with these earlier observations at ${\bm k} = 0$  when compared with the case of 
Fig.~\ref{fig:6}(a) ~\cite{Mondal2021JPCM}.

The effective damping becomes asymmetric across the Brillouin zone when a nonzero chiral inertia 
($C^\prime = 10$~fs) is introduced, as shown in Fig.~\ref{fig:6}(c). The asymmetry between $+k_x$ and $-k_x$
is not restricted to the precessional magnons, but is also evident in the inertial magnons. This behavior clearly 
demonstrates the breaking of reciprocity in magnon dynamics, resulting in direction-dependent dissipation of 
spin waves. Such inertia-induced nonreciprocity provides an additional channel, independent of DMI, for tailoring 
asymmetric magnon transport.

\section{Conclusion}
\label{Sec4}
In this work, we have systematically investigated the role of magnetic inertia in shaping the spin-wave spectrum of a
two-sublattice ferromagnet.
In particular, we investigate the effect of the same sublattice scalar inertia, cross-sublattice scalar inertia, and cross-sublattice chiral inertia on the spin-wave spectrum, magnon group velocity, effective damping, and the effective gyromagnetic ratio.

Our analysis demonstrates that, in the absence of inertia, the magnon bands display 
reciprocity and conventional precessional dynamics without the DMI. The introduction of the same sublattice inertia $\eta$ not only 
modifies the precessional magnon dispersions, but also incorporates the inertial magnon branches.  The same sublattice inertia also reduces the effective damping, thereby offering lower-dissipation channels 
for both precessional and inertial modes. Cross-sublattice inertia $\eta^\prime$ further reshapes the band structure 
by selectively enhancing or reducing the damping of specific modes, while simultaneously shifting the relative 
position of precessional and inertial frequencies. Chiral inertia $C^\prime$ breaks the reciprocity of the spin-wave spectrum 
and induces asymmetry in the group velocity, effective damping, offering a new route to nonreciprocal magnon transport that is independent 
of DMI. More importantly, the introduction of chiral inertia not only influences the inertial branches, but also modifies 
the precessional branches, thereby reshaping the entire magnon spectrum. {   To our knowledge, no {\it ab-initio} calculations of chiral inertia have been reported to date. Nonetheless, CoFeB-based heavy-metal bilayer systems may offer a promising platform for experimental measurement.} 

We have also shown that the crossing between the upper precessional and lower inertial branches can be tuned or 
even eliminated by controlling the strength of the same and cross-sublattice inertial relaxation time, thereby enabling band engineering and gap 
manipulation in magnon spectra. Importantly, our results indicate that inertial magnon modes may possess 
higher group velocities than their precessional counterparts, which may facilitate faster spin-wave information 
transport with a lower energy dissipation channel. These findings demonstrate the crucial role of inertia in magnonic systems and point toward promising strategies 
for harnessing inertia-driven effects in the design of next-generation magnonic devices, offering low-dissipation 
signal transport, tunable nonreciprocity, and enhanced control over spin-wave dynamics.

{ Finally, we provide a symmetry analysis. We note that the presence of a finite magnetization breaks time-reversal symmetry. The presence of a static DMI requires the absence of an inversion symmetry at the bond midpoint. In contrast, the presence of chiral inertia maps onto
an axial vector, which is invariant under inversion and is only forbidden by mirror or other chirality-flipping symmetries. Thus, the symmetries that permit a static DMI are not identical to those that permit chiral inertia. When both DMI and chiral inertia are present, their effects are qualitatively different: DMI contributes to the Hamiltonian and modifies the ground state by yielding a frequency-independent nonreciprocity at low frequencies, whereas chiral inertia modifies the dynamical equation of motion and produces a frequency-dependent nonreciprocity.}

\section{ACKNOWLEDGMENTS}
Financial support by the faculty research scheme at IIT(ISM) Dhanbad, India under Project No. FRS(196)/2023-2024/PHYSICS and Startup Research Grant (SRG) by SERB Under Project No. SRG/2023/000612 is gratefully acknowledged. R.M. acknowledges the Indo-German Science Technology Centre (IGSTC) for the exchange visit via the PECFAR award No. IGSTC/PECFAR/Call 2024/IGSTC-02232/RM-AM/70/2024-25/74.

\section*{Conflict of Interest} 
The authors have no conflicts to disclose. 
  
\section*{Data Availability Statement}
The data that support the findings of this study are available from the corresponding author upon reasonable request.

\appendix
\section{Energy gap dependence on the sublattice magnetic moments and magnetic inertia}
\label{appendixA}
\begin{figure}
    \centering
\includegraphics[scale=0.45]{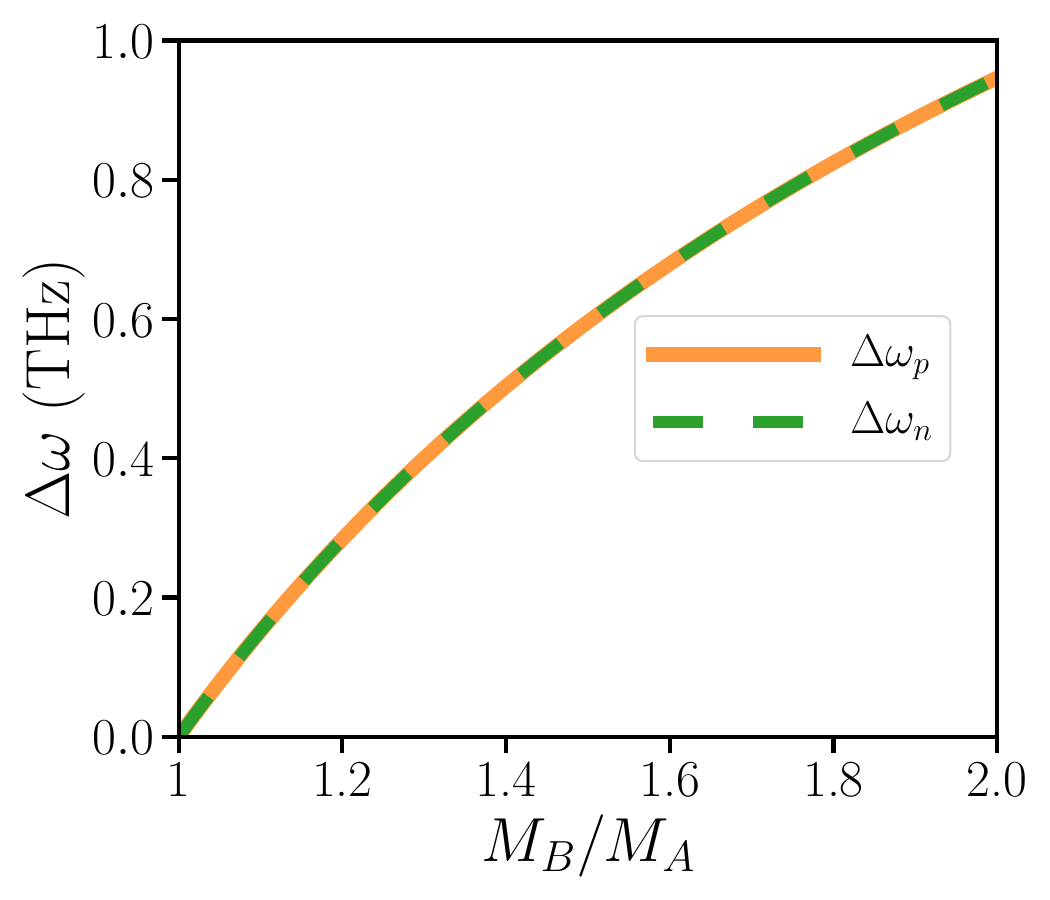}
    \caption{Variation of energy gap at the Brillouin zone edge as a function of sublattice magnetization ratio. Here we used $\eta = 100$ fs, $\eta^\prime = 0$ fs and $C^\prime = 0$ fs.}
    \label{fig:7}
\end{figure}
\begin{figure}[tbh!]
    \centering
\includegraphics[scale=0.45]{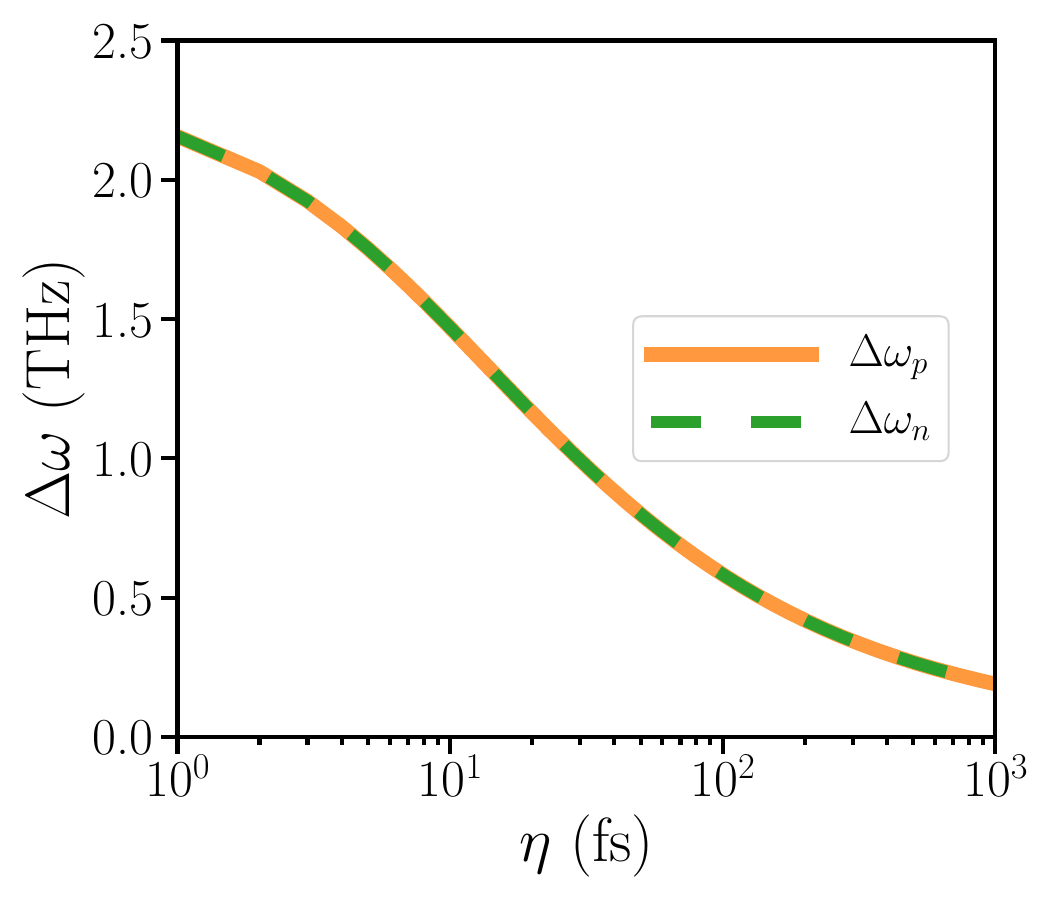}
    \caption{Variation of energy gap at the Brillouin zone edge as a function of $\eta$, setting $\eta^\prime = 0$ and $C^\prime = 0$. We used different magnetization on the two sublattices, viz. $M_A = 1.75 \,\,\mu_B$ and  $M_B = 2.6\,\, \mu_B$.}
    \label{fig:8}
\end{figure}
Due to the asymmetry in sublattice magnetic moments $M_A\neq M_B$, a finite gap opens at the edge of the Brillouin zone, not only between the two precessional modes, but also between the two inertial modes [see Fig. \ref{fig:4}]. We compute the variation of the energy gap as a function of the sublattice magnetization ratio $M_B/M_A$ and isotropic inertial relaxation time $\eta$ in Figs.~\ref{fig:7} and ~\ref{fig:8}, respectively. The energy gap between the two precession modes is defined as 
\begin{align}
    \Delta \omega_p = \omega_{p}^u - \omega_{p}^l \Big\vert_{k_x = \pm \frac{\pi}{a}}\nonumber\\
    \Delta \omega_n = \omega_{n}^u - \omega_{n}^l \Big\vert_{k_x = \pm \frac{\pi}{a}}
\end{align} 

First, we note that for identical magnetic moments $M_A = M_B$, the energy gap at the Brillouin zone edge 
vanishes, as seen in Figs.~\ref{fig:2} and \ref{fig:3}, and is further confirmed in Fig.~\ref{fig:7}. 
In contrast, a difference in sublattice magnetizations opens an energy gap, which increases monotonically 
with the ratio $M_B/M_A$, as illustrated in Fig.~\ref{fig:7}. Interestingly, the energy gaps between the 
two precessional modes, $\Delta \omega_p$, and between the two inertial modes, $\Delta \omega_n$, are found 
to coincide, reflecting a direct correspondence between precessional and inertial branches.

On the other hand, for unequal sublattice magnetizations, the energy gap can be further tuned by the inertial 
relaxation time $\eta$, as demonstrated in Fig.~\ref{fig:8}. For very small values of $\eta$, the energy gap 
is significantly enhanced, whereas increasing $\eta$ progressively reduces the gap. However, a finite gap remains at a large inertial relaxation time $\eta = 1$ ps. Finally, we note that an equal magnetization on the two 
sublattices does not lead to an energy gap opening at the Brillouin zone edge, even upon tuning the inertial relaxation time, $\eta$.

\section{Effect of lower inertial relaxation time on spin-wave spectrum}
\label{appendixB}
\begin{figure*}[tbh!]
    \centering    \includegraphics[width=0.9\linewidth]{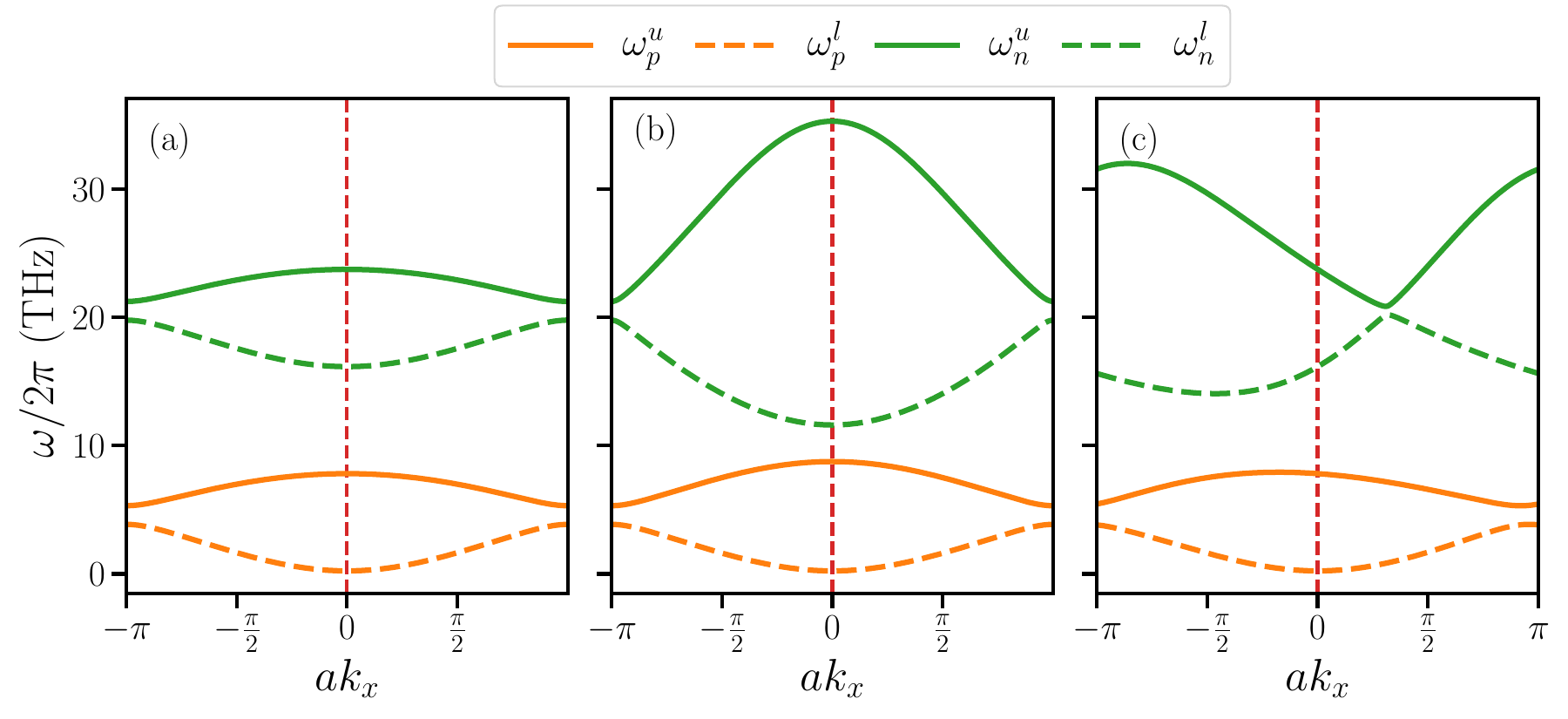}
    \caption{Spin-wave dispersion along $k_x$, setting $k_z = 0$ obtained via solution of Eq. (\ref{Eq12}). The following inertial parameters have been used (a) $\eta = 10$ fs, $\eta^\prime = 0$ fs, $C^\prime = 0$ fs (b) $\eta = 10$ fs, $\eta^\prime = 1$ fs, $C^\prime = 0$ fs, and (c) $\eta = 10$ fs, $\eta^\prime = 0$ fs, $C^\prime = 1$ fs. The other parameters are in Table \ref{tab:1}.}
    \label{fig:9}
\end{figure*}
Here, we compute the same spin-wave spectrum provided in Fig. \ref{fig:4} for a lower inertial relaxation time, $\eta$. In particular, while $\eta = 100$ fs was used in the calculation of Fig. \ref{fig:4}, here we use $\eta = 10$ fs in Fig. \ref{fig:9}. 

We observe that the inertial resonance frequency increases as $1/\eta$, such that the crossing between the 
precessional and inertial branches no longer occurs, as shown in Fig.~\ref{fig:9}(a). These separated magnon 
branches can be further engineered by introducing cross-sublattice inertia $\eta^\prime$ and chiral inertia 
$C^\prime$, as illustrated in Fig.~\ref{fig:9}(b) and \ref{fig:9}(c). For instance, choosing $\eta = 10$~fs and $\eta^\prime = 1$~fs 
reduces the energy gap between the lower inertial mode and the upper precessional mode, while larger values of 
$\eta^\prime$ can drive this gap to vanish entirely. In contrast, a finite chiral inertia ($C^\prime = 1$~fs) induces a stronger asymmetry in the inertial magnon bands compared to the precessional ones, thereby highlighting the 
distinct role of inertia in shaping nonreciprocal magnon spectra.

\bibliographystyle{iopart-num}

\providecommand{\newblock}{}

\end{document}